\documentclass{mem}
\usepackage{natbib}
\usepackage{amsmath}
\usepackage{txfonts}
\usepackage{balance}
\usepackage{flushend}
\usepackage{graphicx}
\usepackage{tikz}
\usepackage{booktabs}
\usepackage[breaklinks,pdftex]{hyperref}
\begin{document}

\title{Clustering analysis of Fermi-LAT unidentified point sources}

\author{G. Cozzolongo\inst{1} \and
        A. M.\,W. Mitchell\inst{1} \and
        S. T. Spencer\inst{1,2} \and
        D. Malyshev\inst{1} \and
        T. Unbehaun\inst{1}
}

\institute{
Friedrich-Alexander-Universität Erlangen-Nürnberg, Erlangen Centre for Astroparticle Physics, 
Nikolaus-Fiebiger-Str. 2, 91058 Erlangen, Germany
\and
Department of Physics, Clarendon Laboratory, Parks Road, Oxford, OX1 3PU, United Kingdom\\
\email{giovanni.cozzolongo@fau.de}
}

\authorrunning{Cozzolongo et al.}
\titlerunning{Clustering of Fermi-LAT Sources}

\date{Received: XX-XX-XXXX (DD-MM-YY); Accepted: XX-XX-XXXX (DD-MM-YY)}

\abstract{
The Fermi Large Area Telescope (LAT) has detected thousands of sources since its launch in 2008, with many remaining unidentified. Some of these point sources may arise from source confusion. Specifically, there could be extended sources erroneously described as groups of point sources. Using the DBSCAN clustering algorithm, we analyze unidentified Fermi-LAT sources alongside some classified objects from the 4FGL-DR4 catalog. We identified 44 distinct clusters containing 106 sources, each including at least one unidentified source. Detailed modeling of selected clusters reveals some cases where extended source models are statistically preferred over multiple point sources. The work is motivated by prior observations of extended TeV gamma-ray sources, such as HESS J1813-178, and their GeV counterparts. In the case of HESS J1813-178, two unidentified Fermi-LAT point sources were detected in the region.
Subsequent multiwavelength analysis combining TeV and GeV data showed that a single extended source is a better description of the emission in this region than two point-like sources.
}

\maketitle

\section{Introduction}
\label{sec:introduction}
The Fermi Gamma-ray Space Telescope, launched on 11th June 2008, is a space-based observatory that has detected thousands of gamma-ray sources. Its primary instrument is the Large Area Telescope (LAT), designed to observe photons in the energy range from 20 MeV to more than 300 GeV. The latest Fermi point source catalog (4FGL-DR4), is based on 14 years of data from 4th August 2008, to 2nd August 2022, and includes 7194 sources detected, of which 81 are spatially extended \citep{2023arXiv230712546B}. 2065 sources in the 4FGL-DR4 catalog remain unclassified, suggesting that some sources may be misclassified due to current analysis limitations. One possibility is that some clusters of point sources may actually be single extended sources, as demonstrated in the case of HESS J1813-178 \citep{2018ApJ...859...69A}. To address this systematically, we have employed the Density-Based Spatial Clustering of Applications with Noise (DBSCAN) algorithm. We apply DBSCAN to the spatial distribution of Fermi-LAT sources, focusing on unassociated sources and those related to usually extended objects. We subsequently perform detailed analyses to determine whether these clusters are better modeled as a collection of point sources or as single extended sources.

\section{Data and Methods}
\label{sec:data_and_methods}
We first perform a clustering analysis of the 4FGL-DR4 catalog sources, and then conduct a detailed morphological and spectral analysis of the identified clusters.

\subsection{Clustering Analysis}
\label{subsec:clustering_analysis}

\begin{figure*}
\centering
\resizebox{0.27\textwidth}{!}{\includegraphics[clip=true]{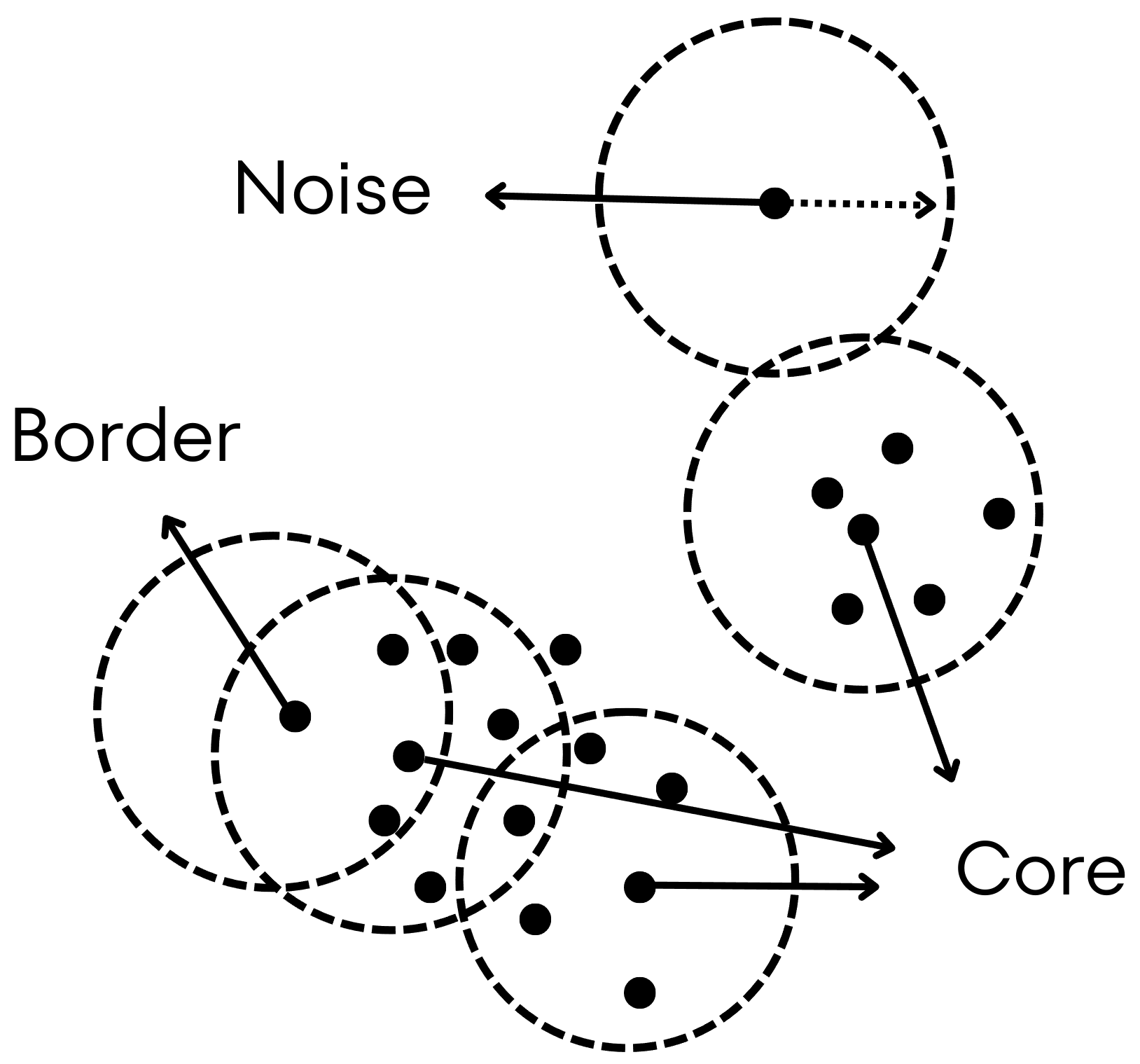}}
\hspace{0.05\textwidth}
\resizebox{0.27\textwidth}{!}{\includegraphics[clip=true]{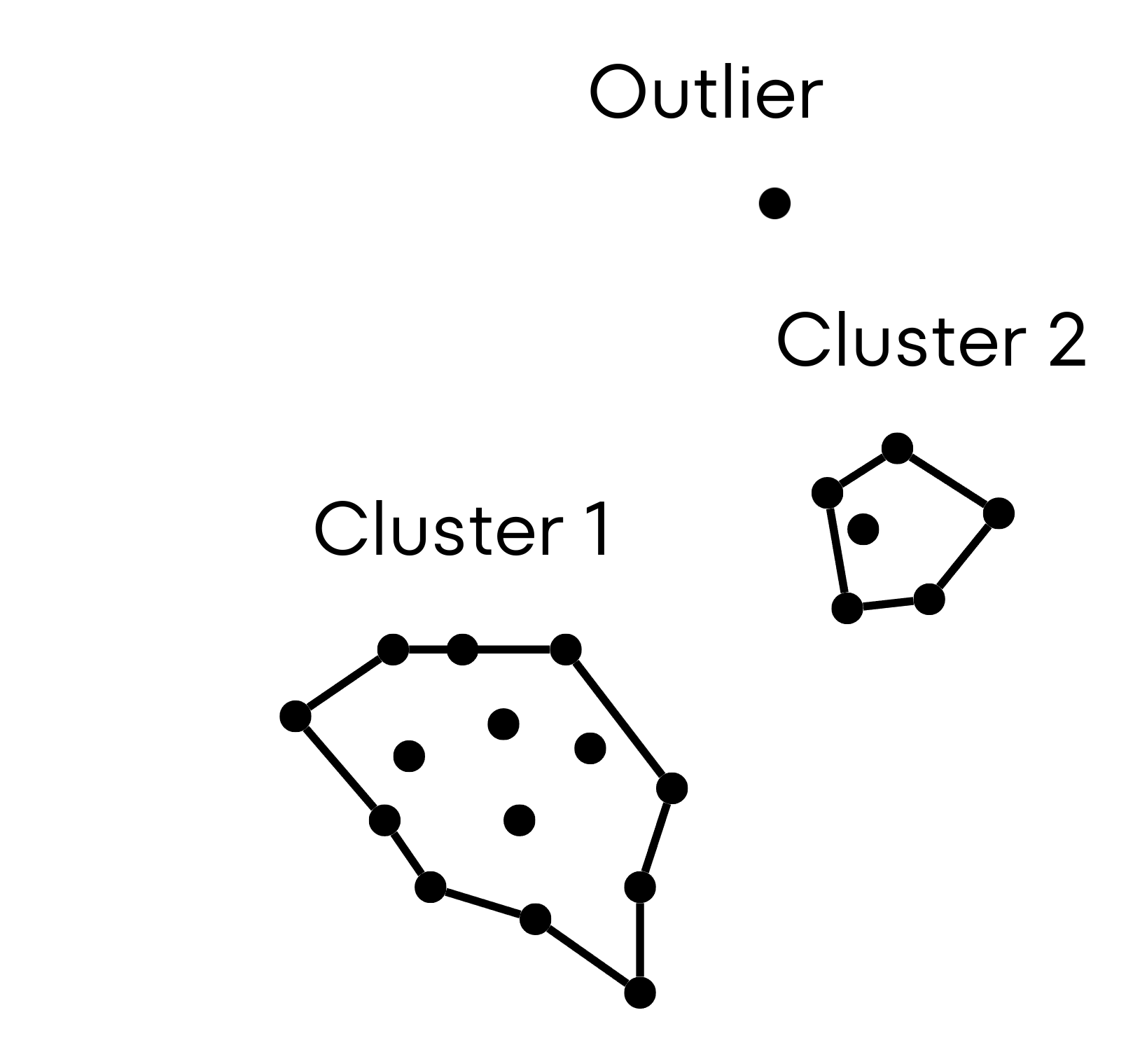}}
\caption{\footnotesize
Three types of points are defined in the DBSCAN algorithm. In this example, two clusters are identified with search radius $1$ and minimum number of points 5.}
\label{fig:dbscan_scheme}
\end{figure*}

The DBSCAN algorithm \citep{1996kddm.conf..226E}, creates a circle around every point and classifies them into core, border or noise points. It operates based on two main parameters: epsilon ($\varepsilon$), defining the maximum distance between two sources for them to be considered neighbors; and MinPts, number of samples required in a neighborhood for a point to be considered a core point (see Fig~\ref{fig:dbscan_scheme}). In our implementation, we set $\varepsilon$ to 0.005 radians (approximately 0.3 degrees) and MinPts to 2. The clustering radius was chosen based on the median radius of the known Fermi-LAT and H.E.S.S. extended sources. %The epsilon value is comparable to the size of the LAT's Point Spread Function (PSF) at high energies, allowing us to identify potentially extended sources that may appear as clusters of point-like emissions.  
We included only unassociated sources and those classified as young pulsars, millisecond pulsars, pulsar wind nebula, supernova remnant, supernova remnant / pulsar wind
nebula, nonblazar active galaxy, or unknown (i.e., unidentified but with a known counterpart in another wavelength) from the 4FGL-DR4 catalog. Our analysis yielded 44 distinct clusters including at least one unidentified source, encompassing a total of 106 sources.

\begin{figure*}
\centering
\resizebox{\hsize}{!}{\includegraphics[clip=true]{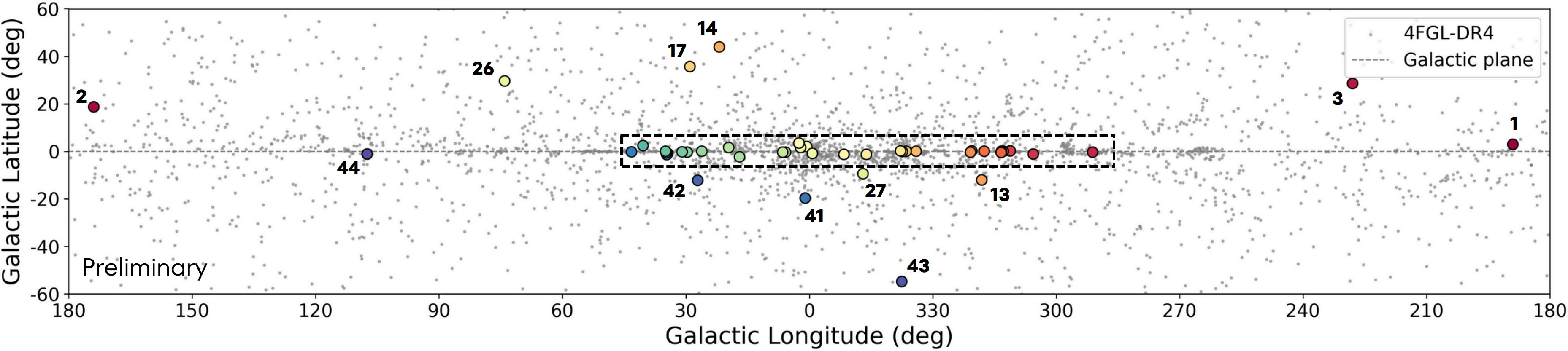}}
\caption{\footnotesize
Fermi-LAT clusters map showing the spatial distribution of the 44 clusters in galactic coordinates. Each cluster is represented by a different color. A detailed picture of the clusters within the black dotted rectangle is provided in Figure~\ref{fig:clusters_hess}.}
\label{fig:clusters_lat}
\end{figure*}

\begin{figure*}
\centering
\resizebox{\hsize}{!}{\includegraphics[clip=true]{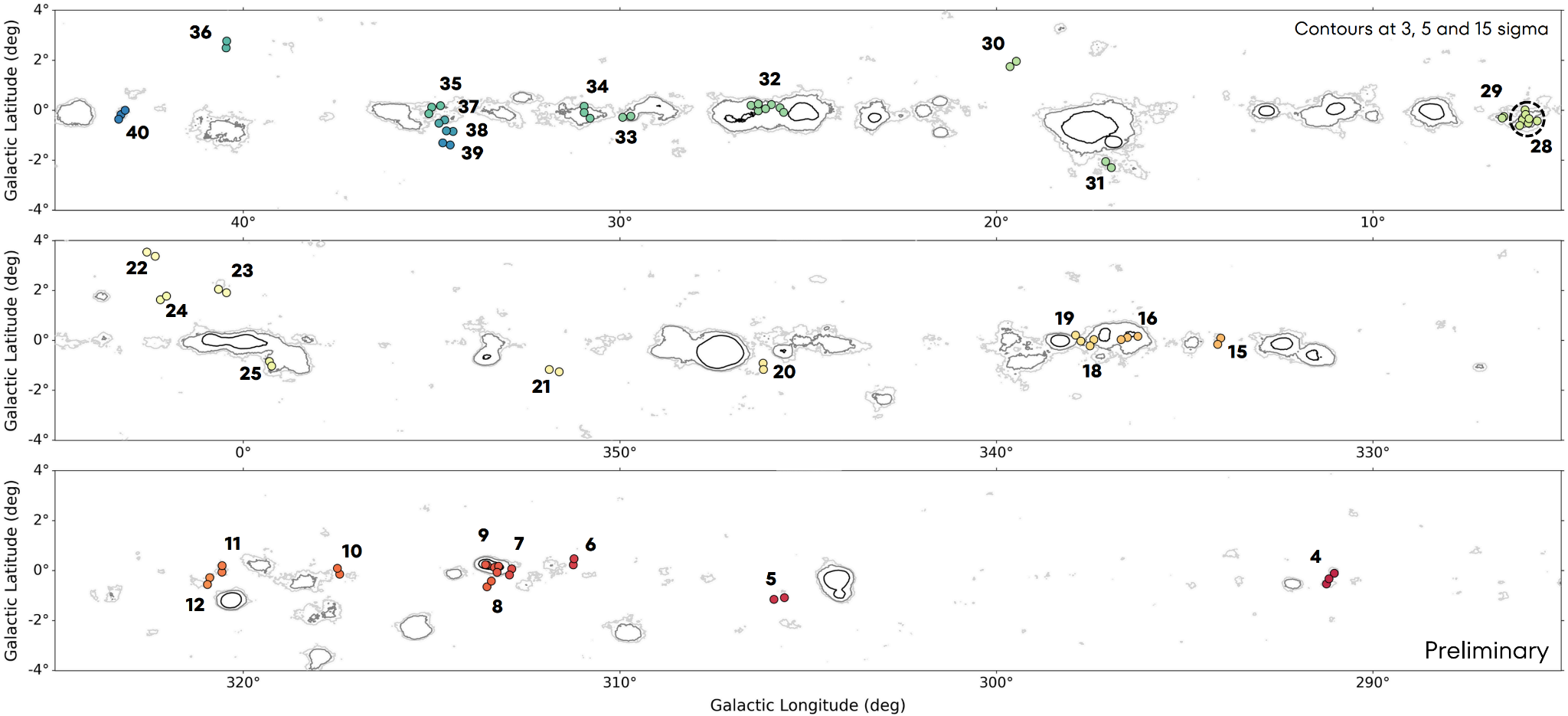}}
\caption{\footnotesize
Our identified clusters with the HGPS contours at 3, 5 and 15 $\sigma$ overlaid, illustrating the spatial coincidence between GeV and TeV gamma-ray sources. Cluster 28, analyzed in detail in the text, is highlighted with a black dotted circle.}
\label{fig:clusters_hess}
\end{figure*}

Fig.~\ref{fig:clusters_lat} presents the spatial distribution of the clusters identified in our analysis. %This map provides a visual representation of the potential extended sources across the Galactic plane, highlighting regions of particular interest for TeV counterpart investigation. 
Fig.~\ref{fig:clusters_hess} shows the clusters overlaid with contours from the H.E.S.S. Galactic Plane Survey (HGPS) \citep{2018A&A...612A...1H}. This comparison allows us to identify potential associations between our GeV clusters and TeV sources.

\subsection{Test criteria for the extension}
\label{subsec:Test_for_extension}
\begin{figure}
\centering
\resizebox{\hsize}{!}{\includegraphics[clip=true]{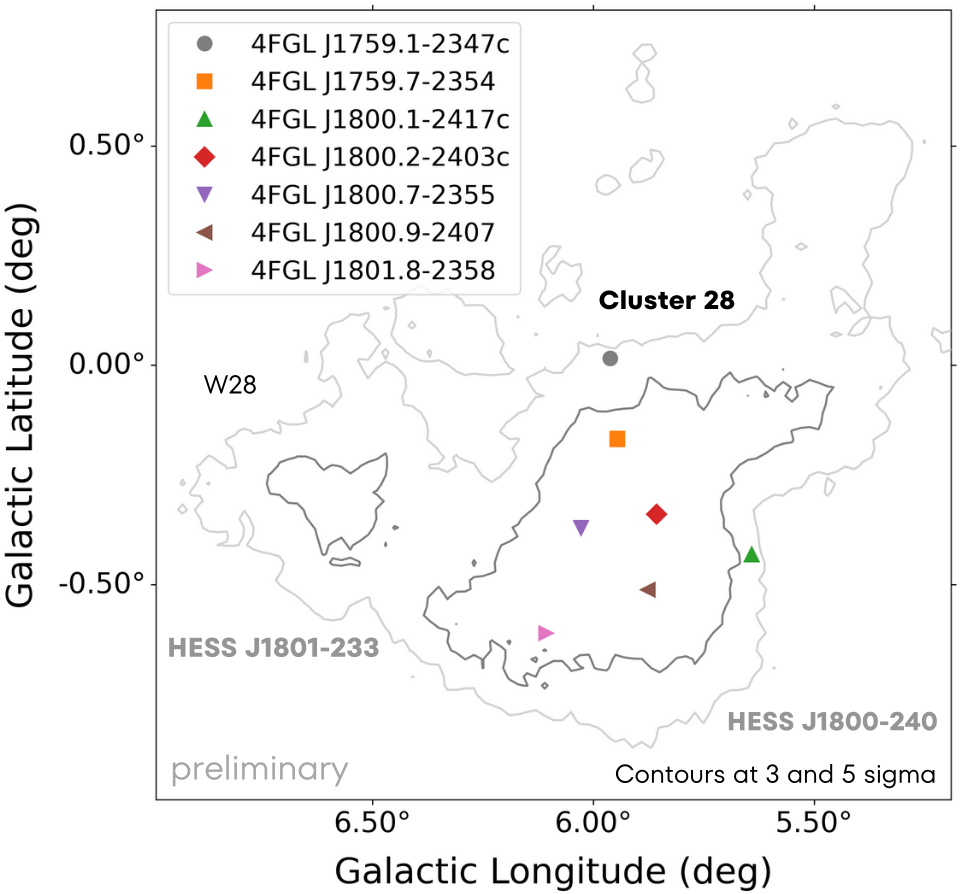}}
\caption{\footnotesize
Cluster 28 H.E.S.S. map. This figure shows a detailed view of the subset of Cluster 28 analyzed in this study, overlaid with H.E.S.S. contours. Contours at 3, 5 and 15 sigma.}
\label{fig:cluster_hess_28}
\end{figure}

Once clusters were identified by DBSCAN, we performed a detailed likelihood analysis of each cluster using the Fermipy v1.2 software package \citep{2017ICRC...35..824W}. For each cluster, we compared the likelihood of the data under two hypotheses: one modeling the emission as multiple point sources, and another modeling it as a single extended source. In the following, we will focus on the analysis of Cluster 28, one of the largest one, as an example (see Figure~\ref{fig:cluster_hess_28}).

We performed a joint likelihood of the PSF event types, which are based on the quality of the reconstructed direction, in the energy range 5 - $10^3$ GeV. We used data collected from 2008 October 27 to 2022 August 1 (the end of the 4FGL-DR4 observational period). For Cluster 28, we reconstructed the events within $6^{\circ}$ of the center of our region of interest (ROI), located at the coordinates $\mathrm{(glat, glon)} = (5.87, -0.51)$.

To quantify the preference for extended source models over multiple point source models, we employed several test statistic definitions and criteria. The likelihood models used in this analysis are defined as follows: $\mathcal{L}_0$ represents the likelihood of the model after removing the clustered sources; $L_\text{ext}$ denotes the likelihood for the extended source model; $L_\text{pt}$ is the likelihood for a single point source model; and ($L_\text{Npts}$) is the likelihood for a model with the original ($N$) point sources.

The TS definitions used were based on the work done by \citet{1996ApJ...461..396M}: the extended source test statistic is defined as $\mathrm{TS} = 2 \ln \left( \mathrm{L_{ext} / L_{0}} \right)$, which measures the significance of detecting an extended source compared to a null hypothesis. The source extension test statistic is given by $\mathrm{TS_{ext}} = 2 \ln \left( \mathrm{L_{ext} / L_{pt}} \right)$, and it quantifies the preference for an extended model over a point source model. Finally, the N-point sources test statistic is $\mathrm{TS_{Npts}} = 2 \ln \left( \mathrm{L_{Npts} / L_{ps}} \right)$, used to compare multiple point source models to the extended source model. Following  \citet{2017ApJ...843..139A}, we claim a detection for sources with a TS $\geq$ 25, which corresponds to $\sim4\sigma$ significance for a single source. To define a source as extended, we used a threshold of $\mathrm{TS_\text{ext}}\geq$ 16, corresponding to nearly $4\sigma$ significance for extension. Given that the significance of non-nested models cannot be quantitatively compared using a simple likelihood ratio test, we considered the Akaike Information Criterion (AIC) \citep{Akaike1974} to determine the preferred model. The AIC is defined as:
\begin{equation}
\mathrm{AIC} = 2k - 2\ln(L)
\end{equation}
where $k$ is the number of free parameters and $L$ is the likelihood.
The definition of this statistical criterion is such that the best available model is the one that minimizes the AIC. Comparing the AIC for extended and point source models leads to:
\begin{equation}
\mathrm{AIC_\text{ext}} < \mathrm{AIC_\text{Npts}} \Rightarrow \mathrm{TS_\text{ext}} > \mathrm{TS_\text{Npts}} - 2\Delta k
\end{equation}
where $\Delta k$ represents the difference in the number of free parameters between the models. The extended source hypothesis was tested using a symmetric disk model and a symmetric Gaussian model, with the radius left as a free parameter in the fit. We chose the model with the highest $\mathrm{TS_{ext}}$. In addition, we performed detailed spectral analyses of the candidate extended sources. We considered simple power laws, log parabolas, and power laws with exponential cutoffs as spectral models. The best-fit spectral model was determined using a likelihood ratio test.

\subsection{Spectral Analysis}
\label{subsec:morphological_analysis}
We performed a binned maximum-likelihood analysis, using eight logarithmic bins per decade in energy and a region of interest (ROI) of $6 \times 6$ degrees with spatial bins of $0.025^\circ$ (as done by \citet{2018ApJS..237...32A} for energies above 1 GeV). After the initial optimization of the ROI, no TS peaks above 25 are left. Next, we optimized all the spectral parameters of the original sources, including the normalization for sources within 3 degrees of the ROI center, the normalization of the isotropic and galactic diffuse models, and the index of the latter. We then optimized the positions of the original sources, together with their spectrum parameters (normalization and index) and all background model parameters. After removing the original sources, we re-optimized the spectral parameters. We then place a point source with a power-law spectrum at the peak of the TS map. We performed a scan over the source width and fit for the extension that maximizes the model likelihood, optimizing also the background model parameters. Finally, we looked for another point source but did not find any with a significance higher than 25.

\section{Results}
\label{sec:results}

Our analysis of the Fermi-LAT data using the DBSCAN algorithm revealed a total of 44 distinct clusters, encompassing a total of 106 individual 4FGL-DR4 sources, with each cluster containing at least one unidentified source. To illustrate our analysis process and results in more detail, we present the case study of one particularly interesting cluster, which we designate as Cluster 28. This cluster includes seven unassociated sources coincident with a TeV source HESS J1800-240. We are studying a subset of the data, excluding 4FGL J1759.1-2347c, for which a point source model is preferred. Specifically, we compared the results between a single extended source model and a model combining the extended source plus one point source. The analysis showed that the latter model is better according to the AIC test, and that point coincided with 4FGL J1759.1-2347c. The TS results for the subset of Cluster 28 are as follows: $\mathrm{TS}$ is 747, $\mathrm{TS_{ext}}$ is 407, and $\mathrm{TS_{6pts}}$ is 378. The difference in the number of degrees of freedom between the extended and point source models is 17.

\begin{figure}
\centering
\resizebox{\hsize}{!}{\includegraphics[clip=true]{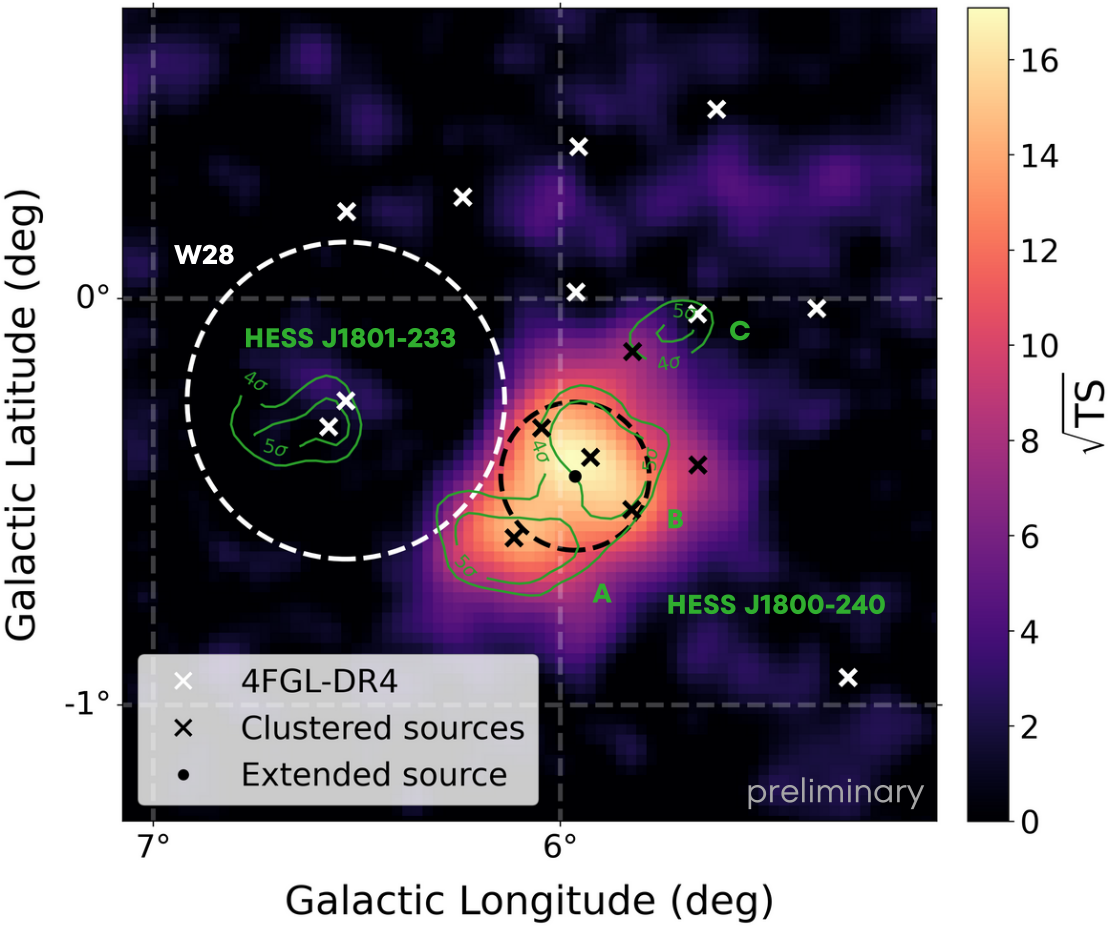}}
\caption{\footnotesize
TS map for Cluster 28. This figure presents the TS map for the subset of Cluster 28 analyzed in this study, highlighting the significance of the emission in this region. The green contours refer to HESS J1801-233 and HESS J1800-240.}
\label{fig:ts_map_28}
\end{figure}

These results indicate that the extended source model is preferred over the multiple point source model for this subset of Cluster 28. The extended model is a Radial Gaussian located at Galactic coordinates $\mathrm{(l, b)} = (5.96\pm0.01, -0.44\pm0.01)\,\mathrm{deg}$, and with radius $\mathrm{r_{68\%}} = (0.26\pm0.01)\,\mathrm{deg}$. The TS map of the region is shown in Figure~\ref{fig:ts_map_28}. Moreover, we performed the comparison of the likelihood of three different spectral models: a power-law, a log-parabola, and a power-law with super-exponential cutoff. We calculated likelihood ratios between these models, with the power-law as the baseline. The extended source is well-described by a power-law model.

\section{Discussion \& Conclusion}
\label{sec:discussion}

The results of our clustering analysis and subsequent detailed modeling provide evidence for the presence of potentially unrecognized extended sources in the Fermi-LAT data. The case of Cluster 28 demonstrates the potential of our approach to reveal complex gamma-ray emitting regions that may not consist of multiple point-sources. The spatial association of the subset of Cluster 28 with the TeV source HESS J1800-240 suggests a physical connection between the GeV and TeV emission. The identification of 44 distinct clusters, encompassing 106 individual sources from the 4FGL-DR4 catalog, provides a valuable starting point for future investigations. Our detailed analysis of a subset of Cluster 28 serves as a prototype for the kind of in-depth study that can be applied to each of these clusters.

Looking ahead, several key areas will be critical for advancing our research. Addressing systematic errors, such as uncertainties in Galactic diffuse emission, the shape of extended sources, and the Fermi-LAT Instrument Response Functions (IRFs). We will also vary the radius in our DBSCAN algorithm to assess the impact on cluster identification. Further investigation into the TeV and multi-wavelength contexts of our identified clusters will provide deeper insights into their nature. This includes conducting joint analyses of individual ROIs using Fermi-LAT and H.E.S.S. data.

\begin{acknowledgements}
GC, AM and STS are supported by the Deutsche Forschungsgemeinschaft (DFG, German Research Foundation) – Project Number 452934793.
\end{acknowledgements}

\bibliographystyle{aa}
\bibliography{references}

\end{document}